%% file: main.tex
\documentclass[conference]{IEEEtran}
\IEEEoverridecommandlockouts
\usepackage{cite}
\usepackage{amsmath,amssymb,amsfonts, amscd,latexsym}
\usepackage{algorithmic}
\usepackage{graphicx}
\usepackage{textcomp}
\usepackage{xcolor}
\usepackage{multirow}
\usepackage{tabularx,multirow,multicol,colortbl,array}
\usepackage{graphicx,color,array}
\usepackage{float}
\usepackage{tikz,tikzscale,bigints,pgfplots}
\usepackage{xcolor}
\usepackage[skip=5pt]{caption}
\usepackage{subcaption}
\usetikzlibrary{positioning,arrows.meta,colorbrewer,calc}
\usepackage{pgfplots}
\pgfplotsset{compat=1.18}

\def\BibTeX{{\rm B\kern-.05em{\sc i\kern-.025em b}\kern-.08em
    T\kern-.1667em\lower.7ex\hbox{E}\kern-.125emX}}
\begin{document}
\title{Analysis of Impulsive Interference in Digital Audio Broadcasting Systems in Electric Vehicles\\
}

\author{\IEEEauthorblockN{1\textsuperscript{st} Chin-Hung Chen}
\IEEEauthorblockA{\textit{Electrical Engineering} \\
\textit{Eindhoven University of Technology}\\
Eindhoven, The Netherlands\\
c.h.chen@tue.nl}
\and
\IEEEauthorblockN{2\textsuperscript{nd} Wen-Hung Huang}
\IEEEauthorblockA{\textit{Electrical Engineering} \\
\textit{Eindhoven University of Technology}\\
Eindhoven, The Netherlands \\
w.huang1@student.tue.nl}
\and
\IEEEauthorblockN{3\textsuperscript{rd} Boris Karanov}
\IEEEauthorblockA{\textit{Electrical Engineering} \\
\textit{Eindhoven University of Technology}\\
Eindhoven, The Netherlands \\
b.p.karanov@tue.nl}
\and
\IEEEauthorblockN{4\textsuperscript{th} Alex Young}
\IEEEauthorblockA{\textit{NXP Semiconductors}\\
	Eindhoven, The Netherlands \\
	alex.young@nxp.com}
\and
\IEEEauthorblockN{5\textsuperscript{th} Yan Wu}
\IEEEauthorblockA{\textit{NXP Semiconductors}\\
	Eindhoven, The Netherlands \\
	yan.wu\_2@nxp.com}
\and
\IEEEauthorblockN{6\textsuperscript{th} Wim van Houtum}
\IEEEauthorblockA{\textit{Electrical Engineering} \\
\textit{Eindhoven University of Technology}\\
\textit{NXP Semiconductors}\\
Eindhoven, The Netherlands \\
wim.van.houtum@nxp.com}
}

\maketitle

\begin{abstract}
    \input{tex/0-Abstract}
\end{abstract}


\section{Introduction} \label{sec:intro}
\input{tex/1-Introduction}

\section{Experiment setup and impulsive noise measurements} \label{sec:data_process}
\input{tex/2-DataProcess}

\section{Markov-Middleton model} \label{sec:mma_model}
\input{tex/3-MMA}

\section{Parameter Estimation} \label{sec:para_esti}
\input{tex/4-ParaEsti}

\section{Optimal Detector Performance}\label{sec:simu}

\input{tex/5-Simu}

\section{Conclusions}\label{sec:summary}
\input{tex/6-Summary}

\section*{Acknowledgment}
\input{tex/ACKs}

\bibliographystyle{IEEEtran}
\bibliography{main}

\vspace{12pt}

\end{document}

%% file: tex/0-Abstract.tex
Recently, new types of interference in electric vehicles (EVs), such as converters switching and/or battery chargers, have been found to degrade the performance of wireless digital transmission systems. Measurements show that such an interference is characterized by impulsive behavior and is widely varying in time. This paper uses recorded data from our EV testbed to analyze the impulsive interference in the digital audio broadcasting band. Moreover, we use our analysis to obtain a corresponding interference model. In particular, we studied the temporal characteristics of the interference and confirmed that its amplitude indeed exhibits an impulsive behavior.
Our results show that impulsive events span successive received signal samples and thus indicate a bursty nature. To this end, we performed a data-driven modification of a well-established model for bursty impulsive interference, the Markov-Middleton model, to produce synthetic noise realization. We investigate the optimal symbol detector design based on the proposed model and show significant performance gains compared to the conventional detector based on the additive white Gaussian noise assumption.

%% file: tex/1-Introduction.tex
\begin{figure*}[ht]
 	\centering
	\begin{subfigure}[b]{\textwidth}
		\centering
		\caption{I component}
		\includegraphics[width=\textwidth]{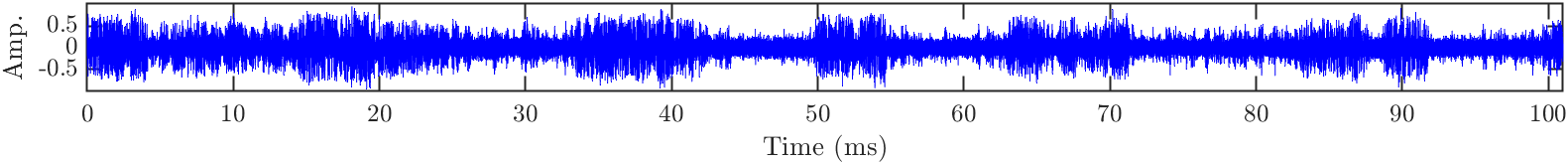}  
		\label{fig:I}
		\vspace*{-5mm} 
	\end{subfigure}
	\begin{subfigure}[b]{\textwidth}
		\centering
		\caption{Q component}
		\includegraphics[width=\textwidth]{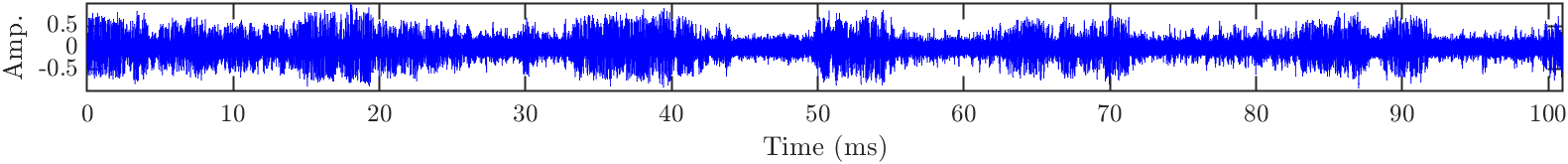}  
		\label{fig:Q} 
		\vspace*{-5mm} 
	\end{subfigure}
	\begin{subfigure}[b]{\textwidth}
		\centering
		\caption{Magnitude}
		\includegraphics[width=\textwidth]{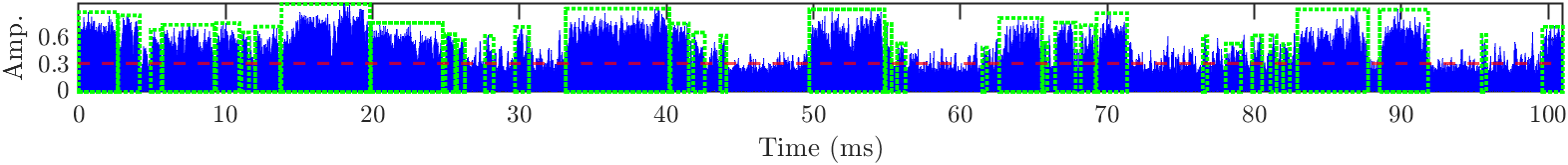}  
		\label{fig:mag}
		\vspace*{-5mm}  
	\end{subfigure}
	\begin{subfigure}[b]{\textwidth}
		\centering
		\caption{Zoomed-in magnitude}
		\includegraphics[width=\textwidth]{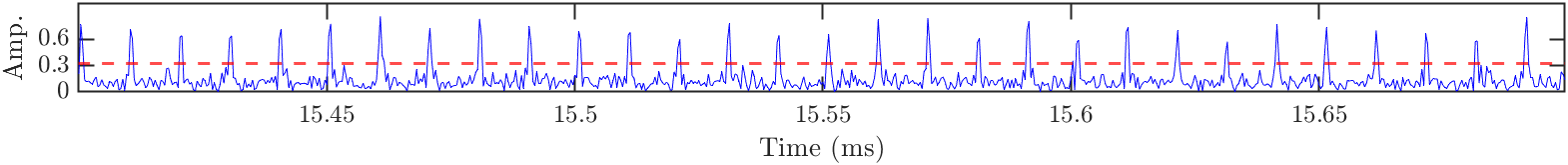}  	
		\label{fig:magzoom}
		\vspace*{-5mm} 
	\end{subfigure}
     \caption{(a) Real part, (b) Imaginary part, (c) magnitude, and (d) roomed in magnitude of the noise and interference measurements from our EV testbed. The green rectangular in (c) represents the captured bursty events with impulsive decision threshold marked with dashed red line}
     \label{fig:IQ} 
\end{figure*}

Electric Vehicles (EVs) represent a significant advancement towards sustainable transportation. However, integrating EVs into the broader mobility landscape introduces new challenges, particularly in the domain of electromagnetic interference~(EMI)~\cite{Kont20, Maou21}. In particular, the power devices in a DC-DC converter are inevitable to generate EMI and cause undesirable interference to the wireless communication systems ~\cite{Gao15, Sakka11}. One critical area affected by EMI is the Digital Audio Broadcasting (DAB) system \cite{Wim22}. The emergence of EVs has underscored the need to understand and mitigate interference in DAB systems to ensure uninterrupted and high-quality broadcasting to these vehicles. To overcome the limitation imposed by the EMI and design an optimal receiver scheme, an accurate interference model is necessary. 

The Middleton impulsive noise model~\cite{Middleton97, Middleton99} is a common tool for EMI analysis. More recently, to describe the bursty nature of the interference, the Markov-Middleton model was proposed in \cite{MMA,Fertonani09}, which introduces a correlation between the impulse events. Moreover, the authors of \cite{MMA} described an optimal symbol detector scheme for such a model based on the Bahl-Cocke-Jelinek-Raviv~(BCJR) algorithm~\cite{bcjr74}.

Nevertheless, the modeling of EV-induced interference has been largely unexplored. Crucially, due to the lack of simple and tractable models, optimal receiver designs for DAB systems have not been thoroughly investigated. In this study, we focused on characterizing via field test recordings and simulating via a modified impulsive interference model, the EMI generated by EVs, particularly emphasizing on the interference encountered in DAB systems. By analyzing the interference signatures, we found that the bursty impulsive EMI behavior is similar to the noise realizations from the Markov-Middleton model. We adopt a data-driven approach to estimate the key parameters for the Markov-Middleton model.

Our work establishes a comprehensive connection between the field-recorded EV interference in the DAB band and a well-established model using statistical analysis and machine learning tools. In more detail, we used the K-means algorithms to cluster the bursty events into background noise and multiple impulsive noise groups. After the algorithm has detected bursty events through an adaptive threshold over the field test data, we estimate the parameters for the modified Markov-Middleton model. Finally, we demonstrate how the inferred parameters can be used to design an optimal receiver for such an interference model.

The paper is organized as follows: Sec.~\ref{sec:data_process} describes the experiment setup and received data characteristics. Sec.~\ref{sec:mma_model} introduces the Markov-Middleton model for communication over bursty impulsive channels. Sec.~\ref{sec:para_esti} details the methods for impulse detection, burst capture, clustering, and model parameter estimation from the measurements. Sec.~\ref{sec:simu} presents a modified Markov-Middleton model and numerical results for such a model. Sec.~\ref{sec:summary} summarize the paper.


%% file: tex/2-DataProcess.tex
\subsection{EV testbed}
We employed a DAB signal recorder mounted on an EV with a dual antenna configuration installed on top of the vehicle to measure the EV interference within the DAB band. This recorder is operated at a $2.6$ MHz sampling rate, where the carrier frequency can be adjusted from $174.928$ MHz to $239.2$ MHz depending on the desired DAB channel. After down-converting the received signal to baseband, the output data from the recorder is in IQ format, which represents the complex number. In our testbed, we also include a Global Positioning System (GPS) and an On-board Diagnostics (OBD) module to record the location and status of the car. These tools assist us in selecting the proper dataset for analyzing EV-induced internal interference.

\subsection{Impulsive noise charateristics}
For illustration purposes, we assume our received signal as 
\begin{equation}\label{eq:yxni}
    Y = X + N + I,
\end{equation}
where $X$ is the transmitted DAB signal, and $Y$ is the noisy received signal from our IQ recorder. Here, we assume $N$ is the thermal noise, which follows a zero mean Gaussian distribution and is independent of the transmitted signal $X$. $I$ represents an additive EV interference. We specifically choose an empty DAB channel ($X=0$) recording to focus on the analysis of the EV interference. 

The time-domain data of our measurement, shown in Fig. ~\ref{fig:IQ}, demonstrate a bursty impulsive behavior, where the sequence of impulsive noise samples is composed of correlated impulse events. This type of impulsive interference causes an elevated noise floor in the frequency domain due to its wideband characteristics, significantly affecting the detection capability of an Orthogonal Frequency Division Multiplexing (OFDM) system such as the DAB receiver. By looking into the consecutive impulsive samples in Fig.~\ref{fig:magzoom}, we found that each impulsive event is separated by a fixed duration of $10$ $\mu$s, likely generated by power electronic devices operated with $100$ kHz.

%% file: tex/3-MMA.tex
This section summarizes the Markov-Middleton model initially proposed for powerline communications ~\cite{MMA}. This model addresses the statistical nature of bursty impulsive noise by incorporating the hidden Markov process into the notable Middleton Class A model. The measurement from our EV testbed shares similar time domain structures with this model, as illustrated in Fig.~\ref{fig:IQ}, offering a promising framework for modeling the bursty behavior of EV interference.

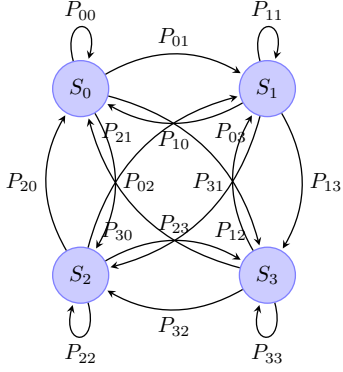
\begin{figure}[ht]
	\centering
	\resizebox{.55\columnwidth}{!}{\input{fig/state_tran.tikz}} 
	\caption{State diagram representation of a 4-state Markov-Middleton model}   
	\label{fig:state_diagram}
\end{figure}

\begin{figure}[hb]
	\centering
	\begin{subfigure}{.24\textwidth}
		\centering
		\caption{I component}
		\includegraphics[width=\textwidth]{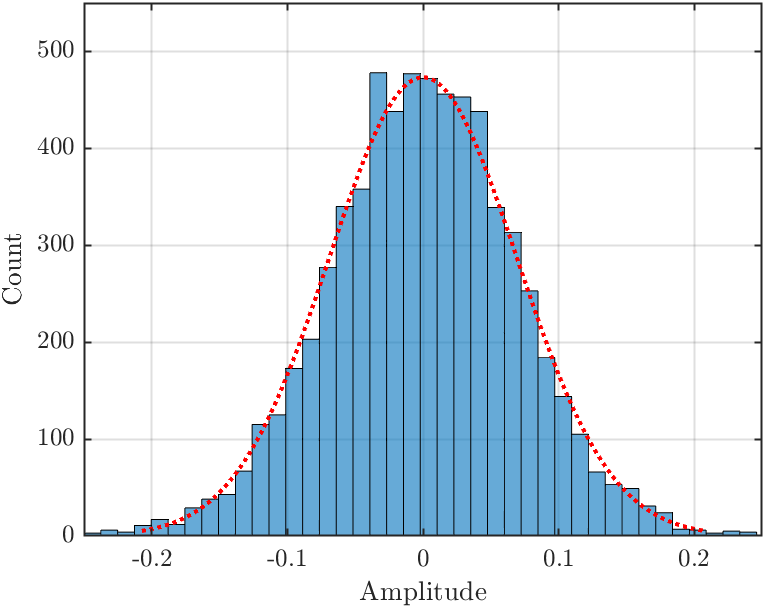}  
		\label{fig:HistI}
		\vspace*{-5mm} 
	\end{subfigure}
	\begin{subfigure}{.24\textwidth}
		\centering
		\caption{Q component}
		\includegraphics[width=\textwidth]{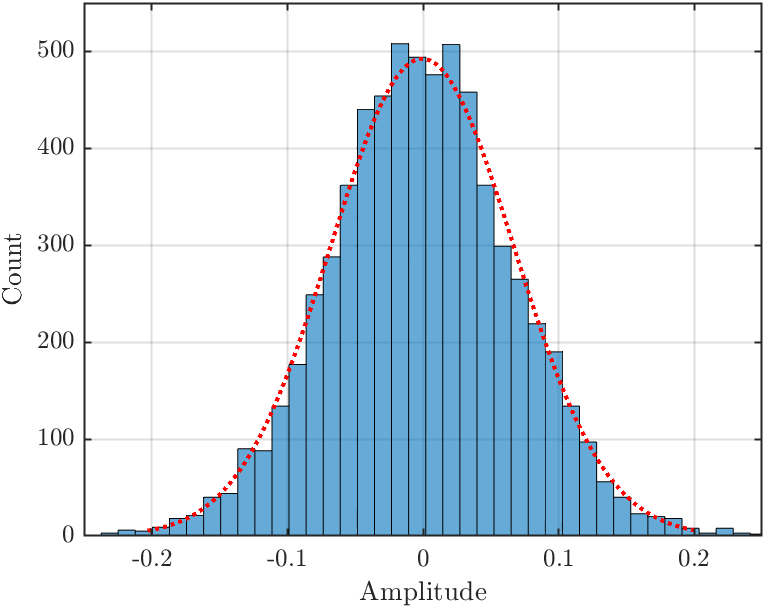}  
		\label{fig:HistQ} 
		\vspace*{-5mm} 
	\end{subfigure}
	\caption{Histograms of (a) I component and (b) Q component of the background noise from the measurements}
	\label{fig:HistIQ} 
\end{figure}

\begin{figure}[bt]
	\centering
	\includegraphics[width=.4\textwidth]{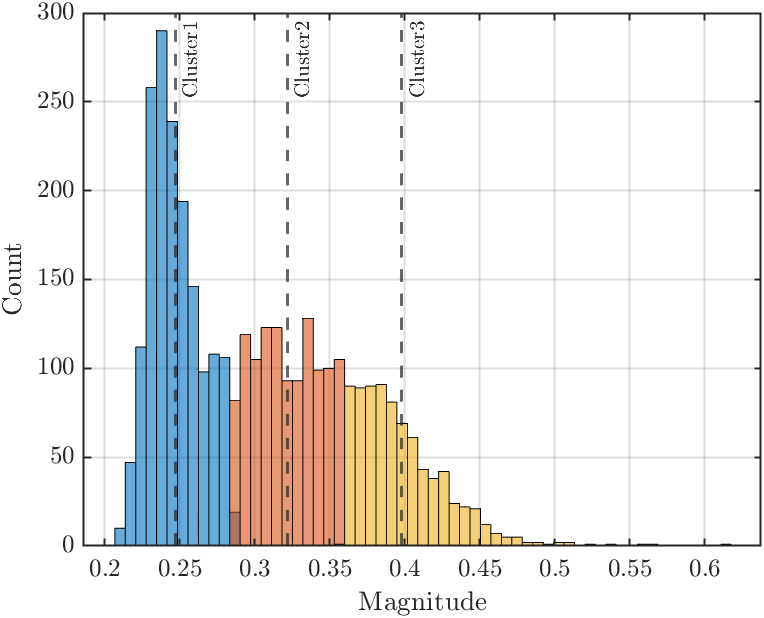} 
	\caption{Histogram of the magnitude of the identified bursty events from the measurements. Three clusters based on the K-mean algorithm with blue, orange, and yellow bars, attributed to State-1, State-2, and State-3, respectively}
	\label{fig:kmean} 
\end{figure}

This model offers a tractable probability density function for the noise sample $z$ as 
\begin{equation}\label{eq:MMApdf}
    p(z) = \sum_{m=0}^{M-1} \frac{p_m}{\sigma_m\sqrt{2\pi}}\exp{\left(-\frac{z^2}{2\sigma^2_m}\right)},
\end{equation}
where 
\begin{equation}\label{eq:pm}
    p_m = \frac{e^{-A}A^m/m!}{\sum_{m=0}^{M-1}e^{-A}A^m/m!}
\end{equation}
and
\begin{equation}\label{eq:varm}
    \sigma^2_m = \sigma^2\frac{m/A+\Gamma}{1+\Gamma},
\end{equation}
with $M$ denoting the total number of noise states, including the background noise. Note that in this model, the noise sample $z$ represents the joint effect of the thermal noise $N$ and interference $I$ in~\eqref{eq:yxni}. The model parameters $A$, $\Gamma$, and $\sigma^2$ represent the impulsive index, the background-to-impulse noise ratio, and the total power of $z$, respectively. Moreover, the Markov-Middleton model includes a parameter $r\in[0, 1]$ that establishes a correlation between noise samples. Then, the state transition matrix can be constructed as 
\begin{equation}\label{eq:P}
    P_{ij} = 
    \begin{cases}
        r + (1-r)p_j, & i=j\\
        (1-r)p_j, & \text{otherwise}.
    \end{cases}
\end{equation}
The state transition diagram for a 4-state noise model is shown in Fig.~\ref{fig:state_diagram}, where the state $S_0$ is denoted as the background noise. Given the state transition probability in~\eqref{eq:P}, we can derive the mean duration (in samples) of each state as
\begin{equation}\label{eq:duration}
    d_m = \frac{1}{(1-r)(1-p_m)}, \quad m = 0,\ldots,M-1.
\end{equation}
As suggested in~\cite{MMA}, the correlation parameter $r$ is derived by measuring the average number of impulse-free consecutive samples. For $m=0$ in~\eqref{eq:duration}, we can derive $r$ as
\begin{equation}\label{eq:r}
	r = 1 - \frac{1}{d_0(1-p_0)}.
\end{equation}

%% file: fig/state_tran.tikz
\tikzstyle{state}=[shape=circle,draw=blue!50,fill=blue!20, minimum size=.9cm]

\begin{tikzpicture}[->,>=stealth,shorten >=1pt,auto,node distance=3cm,
                    semithick]

  \node[state]         (S1)              {$S_0$};
  \node[state]         (S2) [right of=S1] {$S_1$};
  \node[state]         (S3) [below of=S1] {$S_2$};
  \node[state]         (S4) [right of=S3] {$S_3$};

  \path (S1) edge [loop above]    node {$P_{00}$} (S1)
            edge [bend left]     node {$P_{01}$} (S2)
            edge [bend left]     node {$P_{02}$} (S3)
            edge [bend left]     node {$P_{03}$} (S4)
        (S2) edge [bend left]     node {$P_{10}$} (S1)
            edge [loop above]    node {$P_{11}$} (S2)
            edge [bend left]     node {$P_{12}$} (S3)
            edge [bend left]     node {$P_{13}$} (S4)
        (S3) edge [loop below]    node {$P_{22}$} (S3)
            edge [bend left]     node {$P_{20}$} (S1)
            edge [bend left]     node {$P_{21}$} (S2)
            edge [bend left]     node {$P_{23}$} (S4)
        (S4) edge [bend left]     node {$P_{30}$} (S1)
            edge [loop below]    node {$P_{33}$} (S4)
            edge [bend left]     node {$P_{31}$} (S2)
            edge [bend left]     node {$P_{32}$} (S3);
\end{tikzpicture}

%% file: tex/4-ParaEsti.tex
\begin{figure*}[ht]
	\centering
	\includegraphics[width=\textwidth]{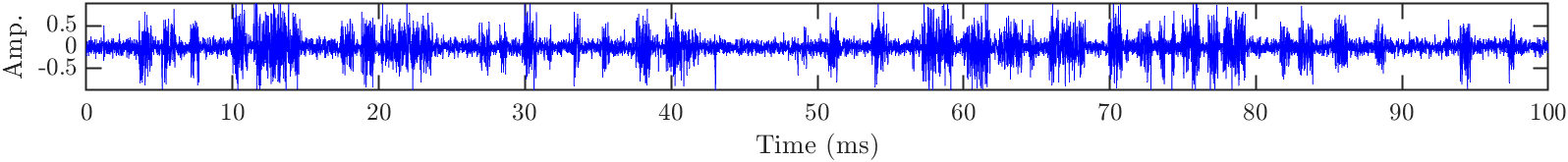} 
	\caption{Time domain synthetic bursty impulsive noise samples generated from the modified Markov-Middleton model with parameters in Table~\ref{tab:exp}}
	\label{fig:modMMA} 
\end{figure*}

The optimal receiver based on the Markov-Middleton model assumption requires accurate parameter estimation. To obtain this, we employed a data-driven approach and selected a representative dataset consisting of bursty events extracted from our measurements. We used this dataset to conduct a comprehensive statistical analysis, which enabled us to explore the connection with the Markov-Middleton model.

\subsection{Impulsive noise detection}
To statistically analyze impulsive noise, we must separate it from background noise. To this end, we establish an adaptive threshold as 
\begin{equation} \label{eq:thr}
    th = \alpha \cdot W_{rms},
\end{equation}
where $W_{rms}$ is the root mean square value of the received signal, and $\alpha$ represents a scaling factor. The effectiveness of the selected threshold value $th$ is assessed by analyzing the histogram of the IQ component of the determined background noise presented in Fig. 3. We observed that the histogram shape conforms reasonably well to a Gaussian distribution.

\begin{table}[tpb]
	\caption{Estimated parameters from the measurements}
	\renewcommand{\arraystretch}{1.3}
	\centering
	\begin{tabular}{|c|c|c|c|c|} 
		\hline
		{\bf State} & {\bf 0} & {\bf 1} &{\bf 2}  & {\bf 3} \\
		\hline
		{\bf $\hat{d}_m$} & 105 & 57 &  69 & 186 \\
		\hline
		{\bf $\hat{p}_m$} & 0.54 & 0.13  & 0.11 & 0.22 \\
		\hline
		{\bf$\hat{\sigma}_m$} & 0.010 & 0.066  & 0.112 & 0.183 \\
		\hline
		{\bf  $\hat{r}$} &  \multicolumn{4}{|c|}{0.979} \\
		\hline
	\end{tabular}
	\label{tab:exp}
\end{table}

\subsection{Bursty event detection}
As was shown in Fig. ~\ref{fig:IQ}, the EV-induced impulsive noise has a bursty nature, where impulses with a similar amplitude appear to occur over a consecutive time interval. In this study, we define a bursty event as a period of time where the signal amplitude surpasses the predefined threshold in~\eqref{eq:thr}. Such events are cataloged through the sequence of indices corresponding to the impulse samples. This study sets the minimum consecutive time for bursty events detection to $0.5$~ms. In addition, the bursty event decision process allows a tolerance margin of $0.3$~$\mu$s for a non-impulsive period. As a result, we detected approximately $3600$ bursty events within an $8$-second recording time. These events are visually denoted by green rectangles in Fig.~\ref{fig:mag}.

\subsection{K-mean clustering}
After identifying all the bursty events in our dataset, we can group them into a set of finite states for further parameter estimation for the Markov-Middleton model. Here, we use the one-dimensional K-mean clustering algorithm to differentiate the bursts into $M-1=3$ impulsive groups based on their average power. The resulting clusters are presented in Fig.~\ref{fig:kmean}, with the centroids of these clusters indicated by black dashed lines. We also observe that the magnitude distribution of the bursty events is similar to the exponential distribution found in \cite{Sacuto14}. After the clustering phase, every bursty event is labeled with a unique cluster ID, effectively differentiating the events based on their mean power. This process facilitates the analysis of impulsive noise in the measurements under the Markov-Middleton model framework by capturing its statistical properties.

\begin{figure}[ht]
	\centering
	\resizebox{.85\columnwidth}{!}{\input{fig/ber_bcjr.tikz}} 
	\caption{BER performance of coded BPSK system with different detectors for the channel with 4-state bursty impulsive noise}   
	\label{fig:ber}
\end{figure}
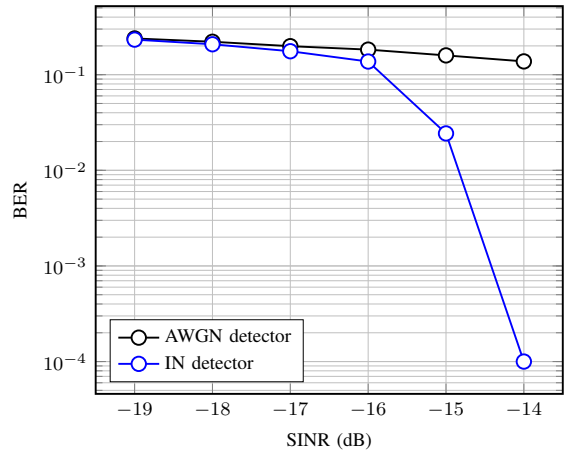

\subsection{Model parameter estimation} \label{sec:MMest}
To infer the probability of occurrence $\hat{p}_m$  for each state from our measurements, we divide the count of bursty events attributed to each state by the total samples of the recording. The estimated mean duration $\hat{d}_m$  of each state is obtained by counting the average number of samples in the bursts. It is important to note that the background noise ($S_0$) exhibits the highest probability of occurrence. However, among the interference states, State-3 ($S_3$) is the most probable, which we found to be different than the Markov-Middleton probability distribution in~\eqref{eq:pm}. This mainly results from the fact that the strongest interference state $S_3$ also had the longest mean duration compared to the other impulsive states. Note that such an observation was also made by the authors of \cite{Sacuto14} regarding their power station measurements, and thus, it's not unusual in real-world EMI scenarios. Given the estimated mean duration $\hat{d}_0$ and state probability $\hat{p}_0$, we can then calculate the correlation parameter $\hat{r}$ using~\eqref{eq:r}. We also compute the sample variance for each state from our measurements. The estimated parameters are listed in Table~\ref{tab:exp}.

%% file: fig/ber_bcjr.tikz
	
\begin{tikzpicture}
	
	\begin{semilogyaxis}[
		axis line style = thick,
		grid = both,
		name = p1,
		font=\footnotesize,
		ylabel = BER, xlabel = SINR (dB),
		legend style={
			font=\footnotesize,
			nodes={scale=1.0},
		},
		legend cell align={left},
		legend pos = south west,
		]
		\addplot[color=black, thick, mark=*, mark options={fill=white}, mark size=3] 
		table[] 
		{
			-19 0.2393
			-18 0.2210
			-17 0.1987
			-16 0.1829
			-15 0.1591
			-14 0.1377    
			
		};
		\addlegendentry{AWGN detector}
		
		\addplot[color=blue, thick, mark=*, mark options={fill=white}, mark size=3] 
		table[] 
		{
			-19 0.2327
			-18 0.2080
			-17 0.1759
			-16 0.1373
			-15 0.0243
			-14 0.0001  
		};
		\addlegendentry{IN detector}
		
	\end{semilogyaxis}

\end{tikzpicture}    

%% file: tex/5-Simu.tex
\subsection{Modified Markov-Middleton model}
Given the parameters inferred from Sec.~\ref{sec:MMest}, we can now use the model to generate the synthetic bursty impulsive noise data. It is important to note that here, we use the data-inferred parameters  $\hat{p}_m$, $\hat{\sigma}_m$, and $\hat{r}$ to compute the probability density function in~\eqref{eq:MMApdf} for each of the impulsive noise states. In contrast, in the original model, the density functions are calculated based on~\eqref{eq:pm} and~\eqref{eq:varm}. The synthetic bursty impulsive noise output sample generated from our modified Markov-Middleton model is shown in Fig.~\ref{fig:modMMA}.

\subsection{Optimal Detector Design}
Based on the modified Markov-Middleton model, an optimal maximum a-posterior detector can be designed to take the interference states into account \cite{MMA,Fertonani09}. This subsection evaluates the bit-error rate for an LDPC-coded binary phase-shift keying ($X \in \{ -1, 1 \}$) system. The codeword length is $64800$, and the code rate is $1/2$. This communication system setting was also used in~\cite{MMA}. Results from Fig.~\ref{fig:ber} show that the IN detector, which incorporates the modified Markov-Middleton noise states, significantly outperforms the conventional AWGN detector that only considers the background noise state.

%% file: tex/6-Summary.tex
In this study, we employed a wideband recorder mounted in an EV to capture the interference plus noise signal in an empty DAB channel. By comprehensively analyzing the measurements, we identified the bursty impulsive nature of the interference and proposed a modified Markov-Middleton model with the parameters estimated from our recordings. This research contributes to providing a systematic analysis of EV-induced interference, thereby facilitating the development of robust receiver design. This study aims to bridge the gap between simple theoretical models and the real-world phenomena of EV interference.

%% file: tex/ACKs.tex
This work was funded by the RAISE collaboration framework between Eindhoven University of Technology and NXP, including a PPS-supplement from the Dutch Ministry of Economic Affairs and Climate Policy

%% file: main.bbl
\begin{thebibliography}{10}
\providecommand{\url}[1]{#1}
\csname url@samestyle\endcsname
\providecommand{\newblock}{\relax}
\providecommand{\bibinfo}[2]{#2}
\providecommand{\BIBentrySTDinterwordspacing}{\spaceskip=0pt\relax}
\providecommand{\BIBentryALTinterwordstretchfactor}{4}
\providecommand{\BIBentryALTinterwordspacing}{\spaceskip=\fontdimen2\font plus
\BIBentryALTinterwordstretchfactor\fontdimen3\font minus
  \fontdimen4\font\relax}
\providecommand{\BIBforeignlanguage}[2]{{%
\expandafter\ifx\csname l@#1\endcsname\relax
\typeout{** WARNING: IEEEtran.bst: No hyphenation pattern has been}%
\typeout{** loaded for the language `#1'. Using the pattern for}%
\typeout{** the default language instead.}%
\else
\language=\csname l@#1\endcsname
\fi
#2}}
\providecommand{\BIBdecl}{\relax}
\BIBdecl

\bibitem{Kont20}
K.~Pliakostathis, ``Research on {EMI} from modern electric vehicles and their
  recharging systems,'' in \emph{2020 International Symposium on
  Electromagnetic Compatibility - EMC EUROPE}, 2020, pp. 1--6.

\bibitem{Maou21}
A.~Maouloud, M.~Klingler, and P.~Besnier, ``A test setup to assess the impact
  of {EMI} produced by on-board electronics on the quality of radio reception
  in vehicles,'' \emph{IEEE Transactions on Electromagnetic Compatibility},
  vol.~63, no.~6, pp. 1844--1855, 2021.

\bibitem{Gao15}
X.~Gao, D.~Su, and Y.~Li, ``Study on electromagnetic interference of {DC/DC}
  converter used in the {EV},'' in \emph{2015 Asia-Pacific Symposium on
  Electromagnetic Compatibility (APEMC)}, 2015, pp. 258--261.

\bibitem{Sakka11}
\BIBentryALTinterwordspacing
M.~A. Sakka, J.~V. Mierlo, and H.~Gualous, ``{DC/DC} converters for electric
  vehicles,'' in \emph{Electric Vehicles}, S.~Soylu, Ed.\hskip 1em plus 0.5em
  minus 0.4em\relax Rijeka: IntechOpen, 2011, ch.~13. [Online]. Available:
  \url{https://doi.org/10.5772/17048}
\BIBentrySTDinterwordspacing

\bibitem{Wim22}
W.~J. van Houtum, ``Time-division spatial interference rejection ({TDSIR})-
  procedure.''\hskip 1em plus 0.5em minus 0.4em\relax U.S. Patent 11722197B2,
  2022.

\bibitem{Middleton97}
D.~Middleton, ``Statistical-physical models of electromagnetic interference,''
  \emph{IEEE Transactions on Electromagnetic Compatibility}, vol. EMC-19,
  no.~3, pp. 106--127, 1977.

\bibitem{Middleton99}
------, ``Non-{G}aussian noise models in signal processing for
  telecommunications: new methods an results for class {A} and class {B} noise
  models,'' \emph{IEEE Transactions on Information Theory}, vol.~45, no.~4, pp.
  1129--1149, 1999.

\bibitem{MMA}
G.~Ndo, F.~Labeau, and M.~Kassouf, ``A {M}arkov-{M}iddleton model for bursty
  impulsive noise: Modeling and receiver design,'' \emph{IEEE Transactions on
  Power Delivery}, vol.~28, no.~4, pp. 2317--2325, 2013.

\bibitem{Fertonani09}
D.~Fertonani and G.~Colavolpe, ``On reliable communications over channels
  impaired by bursty impulse noise,'' \emph{IEEE Transactions on
  Communications}, vol.~57, no.~7, pp. 2024--2030, 2009.

\bibitem{bcjr74}
L.~Bahl, J.~Cocke, F.~Jelinek, and J.~Raviv, ``Optimal decoding of linear codes
  for minimizing symbol error rate (corresp.),'' \emph{IEEE Transactions on
  Information Theory}, vol.~20, no.~2, 1974.

\bibitem{Sacuto14}
F.~Sacuto, F.~Labeau, and B.~L. Agba, ``Wide band time-correlated model for
  wireless communications under impulsive noise within power substation,''
  \emph{IEEE Transactions on Wireless Communications}, vol.~13, no.~3, pp.
  1449--1461, 2014.

\end{thebibliography}
